\documentclass[english,superscriptaddress,aps,prb,reprint]{revtex4-1}
\usepackage{mathptmx}
\usepackage{helvet}
\usepackage[T1]{fontenc}
\usepackage[latin9]{inputenc}
\usepackage{color}
\usepackage{babel}
\usepackage{amsmath}
\usepackage{amssymb}
\usepackage{tabularx}
\usepackage{colortbl,booktabs}
\usepackage{textcomp}
\usepackage[compact]{titlesec}
\usepackage[unicode=true,pdfusetitle,
bookmarks=true,bookmarksnumbered=false,bookmarksopen=false,
breaklinks=false,pdfborder={0 0 1},backref=false,colorlinks=true]{hyperref}
\hypersetup{linkcolor=blue,citecolor=blue}

\makeatletter

\usepackage{tikz}
\usetikzlibrary{external}
\tikzsetexternalprefix{figures/}
\tikzset{font={\fontsize{10pt}{12}\selectfont}}

\usepackage[american,arrowmos]{circuitikz}
\tikzset{x=0.6cm,y=0.6cm}
\ctikzset{bipoles/length=0.7cm}

\usepackage{pgfplots}
\usepgfplotslibrary{groupplots}
\usepgfplotslibrary{statistics}
\pgfplotsset{compat=1.3}
\usetikzlibrary{arrows.meta}
\tikzset{>=Latex}

\newcommand{\yline}[1]{
	\draw[dashed] ({rel axis cs:0,0} |- {#1}) -- ({rel axis cs:1,0} |- {#1})
}

\newcommand{\xshade}[3]{
	\draw [fill=#3, fill opacity=0.1, draw=none]
	({rel axis cs:0,0} -| {#1}) rectangle
	({rel axis cs:0,1} -| {#2})
}
\newcommand{\yshade}[3]{
	\draw [fill=#3, fill opacity=0.1, draw=none]
	({rel axis cs:0,0} |- {#1}) rectangle
	({rel axis cs:1,0} |- {#2})
}
\definecolor{redorange}{HTML}{C84127}
\definecolor{cedar}{HTML}{3D2117}
\definecolor{maroon}{HTML}{6F2B2B}
\definecolor{moss}{HTML}{896E44}
\definecolor{beige}{HTML}{FEFCE8}

\newcommand{\plotdatadir}{plotdata}

\makeatother

\begin{document}

\title{Stochastic IMT (insulator-metal-transition) neurons: An interplay
	of thermal and threshold noise at bifurcation}

\author{Abhinav Parihar}

\affiliation{Georgia Institute of Technology, Altanta, GA}

\author{Matthew Jerry}

\affiliation{University of Notre Dame, Notre Dame, IN}

\author{Suman Datta}

\affiliation{University of Notre Dame, Notre Dame, IN}

\author{Arijit Raychowdhury}

\affiliation{Georgia Institute of Technology, Altanta, GA}

\begin{abstract} 
	
Artificial neural networks can harness stochasticity in multiple ways to enable
a vast class of computationally powerful models. Boltzmann machines and other
stochastic neural networks have been shown to outperform their deterministic
counterparts by allowing dynamical systems to escape local energy minima.
Electronic implementation of such stochastic networks is currently limited to
addition of algorithmic noise to digital machines which is inherently
inefficient; albeit recent efforts to harness physical noise in devices for
stochasticity have shown promise. To succeed in fabricating electronic
neuromorphic networks we need experimental evidence of devices with measurable
and controllable stochasticity which is complemented with the development of
reliable statistical models of such observed stochasticity. Current research
literature has sparse evidence of the former and a complete lack of the latter.
This motivates the current article where we demonstrate a stochastic neuron
using an insulator-metal-transition (IMT) device, based on electrically induced
phase-transition, in series with a tunable resistance. We show that an IMT
neuron has dynamics similar to a piecewise linear FitzHugh-Nagumo (FHN) neuron
and incorporates all characteristics of a spiking neuron in the device
phenomena. We experimentally demonstrate spontaneous stochastic spiking along
with electrically controllable firing probabilities using Vanadium Dioxide (VO$_2$)
based IMT neurons which show a sigmoid-like transfer function. The stochastic
spiking is explained by two noise sources - thermal noise and threshold
fluctuations, which act as precursors of bifurcation. As such, the IMT neuron is
modeled as an Ornstein-Uhlenbeck (OU) process with a fluctuating boundary
resulting in transfer curves that closely match experiments. The moments of
interspike intervals are calculated analytically by extending the
first-passage-time (FPT) models for Ornstein-Uhlenbeck (OU) process to include a
fluctuating boundary. We find that the coefficient of variation of interspike
intervals depend on the relative proportion of thermal and threshold noise,
where threshold noise is the dominant source in the current experimental
demonstrations. As one of the first comprehensive studies of a stochastic neuron
hardware and its statistical properties, this article would enable efficient
implementation of a large class of neuro-mimetic networks and algorithms.

	
\end{abstract}

\maketitle

\section{Introduction}

A growing need for efficient machine-learning in autonomous systems coupled with
an interest in solving computationally hard optimization problems has led to
active research in stochastic models of computing. Optimization techniques
\citep{haykin_neural_2009-1} including Stochastic Sampling Machines (SSM),
Simulated Annealing, Stochastic Gradients etc. are examples of such models. All
these algorithms are currently implemented using digital hardware which first
creates a mathematically accurate platform for computing, and later adds digital
noise at the algorithm level. Hence, it is enticing to construct hardware
primitives that can harness the already existing physical sources of noise to
create a stochastic computing platform. The principal challenge with such
efforts is the lack of stable or reproducible distributions, or functions of
distributions, of physical noise. One basic stochastic unit which enables a
systematic construction of stochastic hardware has long been known - the
stochastic neuron \citep{gerstner_spiking_2002} - which is also believed to be
the unit of computation in the  human brain. Moreover, recent studies
\citep{buesing_neural_2011} have demonstrated practical applications like
sampling using networks of such stochastic spiking neurons. There have been some
attempts for building neuron hardware \citep{mehonic2016emulating,
	sengupta2016magnetic, tuma2016stochastic, pickett2013scalable,
	indiveri2006vlsi}, but building a neuron with self-sustained spikes, or
oscillations, which are stochastic in nature and where the probability of firing
is controllable using a signal has been challenging. Here, we demonstrate and
analytically study a true stochastic neuron \citep{Jerry2017} which is
fabricated using oscillators \citep{shukla_synchronized_2014,
	shukla_pairwise_2014, parihar_synchronization_2015} based on insulator-metal
transition (IMT) materials, e.g. Vanadium Dioxide (VO$_{2}$), wherein the
inherent physical noise in the dynamics is used to implement stochasticity. The
firing probability, and not just the deterministic frequency of oscillations or
spikes, is controllable using an electrical signal. We also show that such an
IMT neuron has similar dynamics as a piecewise linear FitzHugh-Nagumo (FHN)
neuron with thermal noise along with threshold fluctuations as precursors of
bifurcation resulting in a sigmoid-like transfer function for the neural firing
rates. By analyzing the variance of interspike interval, we determine that for
the range of thermal noise present in our experimental demonstrations, threshold
fluctuations are responsible for most of the stochasticity compared to thermal
noise.

\section{Materials and Methods}

\subsection{IMT phase change neuron model}

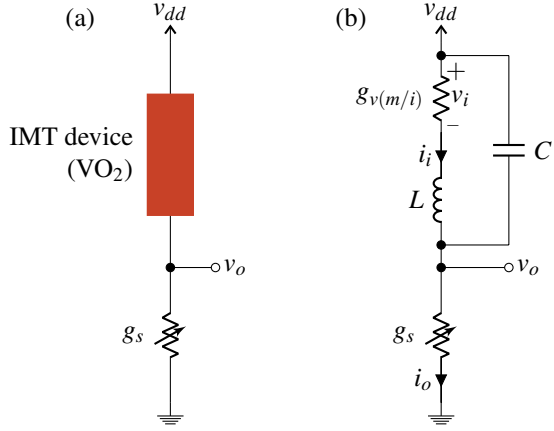
\begin{figure}
	\centering
	\tikzsetnextfilename{vo2_circuit}
	\begin{tikzpicture}
	\renewcommand{\baselinestretch}{1}
	\draw
		(0,0) node[ground]{} 
		to[vR=$g_s$,invert] ++(0,3)
		to[/tikz/circuitikz/bipoles/length=2cm,fullgeneric,color=redorange]
		++(0,5) node(vdd)[vdd]{$v_{dd}$}
		++(0,-5) to[short,*-o] ++(1,0) node[anchor=west]{$v_o$};
		\node at (-2,8.5) {(a)};
		\node[align=right] at (-2.2,5.5) {IMT device\\(VO$_{2}$)};
	
	\draw
		(6,0) node[ground]{}
		to[vR=$g_s$,invert,i<^=$i_{o}$] ++(0,3)
		to[L=$L$,i^<=$i_{i}$] ++(0,3)
		to[R=$g_{v(m/i)}$,v<=$v_{i}$] ++(0,1.5) 
		-- ++(0,0.5) node[vdd]{$v_{dd}$}
		
		++(0,-4.5) to[short,*-] ++(1.5,0)
		to[C,l_=$C$] ++(0,4.2) to[short,-*] ++(-1.5,0) ++(0,-4.7)
		to[short,*-o] ++(1.5,0) node[anchor=west]{$v_o$};
		\node at (4,8.5) {(b)};
	
\end{tikzpicture}
	\caption{(a) VO$_{2}$ based IMT spiking neuron circuit consisting of a VO$_{2}$
		device in series with a tunable resistance. (b) Equivalent circuit
		of IMT neuron using a series inductance $L$ and a parallel capacitance $C$}
	\label{fig:vo2_circuit}
\end{figure}

A stochastic IMT neuron is fabricated using relaxation oscillators
\citep{shukla_synchronized_2014,parihar_synchronization_2015} composed of an IMT
phase change device, e.g. Vanadium Dioxide (VO$_{2}$), in series with a tunable
resistance, e.g. transistor \citep{shukla_pairwise_2014} (Figure
\ref{fig:vo2_circuit}a). An IMT device is a two terminal device with two
resistive states - insulating (I) and metallic (M), and the device transitions
between the two states based on the applied electric field (which in turn
changes the current through the device and the corresponding temperature) across
it. The phase transitions are hysteretic in nature, which means that the IMT
(insulator-to-metal) transition does not occur at the same voltage as the MIT
(metal-to-insulator) transition. For a range of values of the series resistance,
the resultant circuit shows spontaneous oscillations due to hysteresis and a
lack of stable point \citep{parihar_synchronization_2015}. Overall, the series
resistance acts as a parameter for bifurcation between a spiking (or
oscillating) state and a resting state of an IMT neuron.

The equivalent circuit model for an IMT oscillator is shown in Figure
\ref{fig:vo2_circuit}b with the hysteretic switching conductance
$g_{v(m/i)}$ ($g_{vm}$ in metallic and $g_{vi}$ in insulating state),
a series inductance $L$, and a parallel internal capacitance $C$.
Let the IMT and MIT thresholds of the device be denoted by $v_{h}$
and $v_{l}$ respectively, with $v_{h}>v_{l}$, and the current-voltage
relationship of the hysteretic conductance be
\[
v_{i}=h(i_{i},s)
\]
where $h$ is linear in $i_i$ and $s$ is the state - metallic (M)
or insulating (I).

The system dynamics is then given by:
\begin{eqnarray}
	L\frac{di_{i}}{dt} & = & (v_{dd}-h(i_{i},s))-v_{o}\nonumber \\
	C\frac{dv_{o}}{dt} & = & i_{i}-g_{s}v_{o}\label{eq:vo2_system}
\end{eqnarray}
with $i_{i}$ and $v_{o}$ as shown in figure \ref{fig:vo2_circuit}b and $s$ is 
considered as an independent variable.

\subsection{Mechanism of oscillations and spikes}

In VO$_{2}$, IMT and MIT transitions are orders of magnitude faster than RC time
constants for oscillations, as observed in frequency  \citep{kar_intrinsic_2013}
and time-domain measurements for voltage driven \citep{jerry2016dynamics} and
photoinduced transitions  \citep{cocker2012phase}. As such, the change in
resistance of the IMT device is assumed to be instantaneous. Figure
\ref{fig:imt_traj}a shows the phase space $i_{i} \times (v_{dd}-v_{o})$. V-I
curves for IMT device in the two states metallic (M) and insulating (I) and the
load line for series conductance  $v_{o}=i_{i}/g_{s}$ for the steady state are
shown along with the fixed points of the system $S_{1}$ and $S_{2}$ in
insulating and metallic states respectively. The load line and V-I curves are
essentially the nullclines of $v_{o}$ and $i_{i}$ respectively. The capacitance-
inductance pair delays the transitions and slowly pulls the system towards the
fixed points S$_{1}$ and S$_{2}$ even when the IMT device transitions
instantaneously. For small $L/C$ ratio, the eigenvector (of the coefficient matrix) 
with large negative
eigenvalue becomes parallel to the x-axis, whereas the other eigenvector becomes
parallel to AB' or BA' depending on the state (M or I). When the system
approaches A from below (or B from above) and IMT device is insulating (or
metallic) with fixed point $S_{1}$ (or $S_{2}$), the IMT device transitions into
metallic (or insulating) state changing the fixed point to S$_{2}$ (or $S_{1}$).
Two trajectories are shown starting from points A and B each for the system
(\ref{eq:vo2_system}) - one for small $L/C$ value (solid) and the other for
large $L/C$ value (dashed). After a transition, the system moves parallel to
$x$-axis almost instantaneously and spends most of the time following the V-I
curve towards the fixed point. Before the fixed point is reached the MIT (or
IMT) transition threshold is encountered which switches the fixed point, and the
cycle continues resulting in sustained oscillations or spike generation.

\subsection{Model approximations and connections with FHN neuron}

\subsubsection{Non-hysteretic approximation}

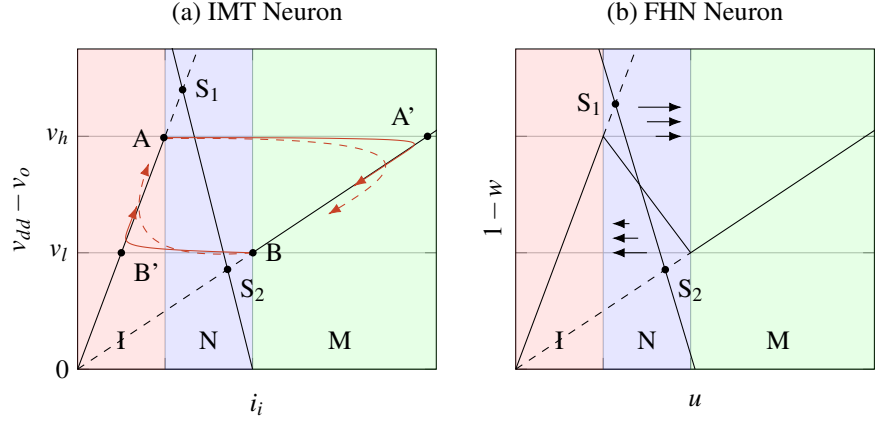
\begin{figure*}
	\centering
	\tikzsetnextfilename{imt_traj}
	\begin{tikzpicture}
\renewcommand{\baselinestretch}{1}
\begin{groupplot}[
	group style={
		group size=2 by 1,
		xlabels at=edge bottom,
		xticklabels at=edge bottom,
		ylabels at=edge left,
		yticklabels at=edge left,
		horizontal sep=30pt,
	},
	xmin=0,
	xmax=0.82,
	ymin=0,
	ymax=1.1,
	ytick={0,0.4,0.8},
	yticklabels={0,$v_l$,$v_h$},
	xtick={0.2,0.4},
	xticklabels=\empty,
    width=2.5in,
	height=2.3in,
	grid=major
]

\nextgroupplot[
	xlabel={$i_{i}$},
	ylabel={$v_{dd}-v_{o}$},
	title=(a) IMT Neuron]
	\xshade{axis cs:0,0}{axis cs:0.2,0}{red};
	\xshade{axis cs:0.2,0}{axis cs:0.4,0}{blue};
	\xshade{axis cs:0.4,0}{axis cs:1.1,0}{green};
	%\yshade{axis cs:0,0.4}{axis cs:0,0.8}{teal};
	\addplot [domain=0:0.2] {4*x}
		coordinate [pos=0.5] (l2)
		coordinate [pos=1] (h1);
	\addplot [domain=0.4:1.2] {x}
		coordinate [pos=0] (l1)
		coordinate [pos=0.5] (h2);
	\addplot [domain=0.2:0.3,dashed] {4*x};
	\addplot [domain=0:0.4,dashed] {x};
	\addplot [domain=0.2:0.4] {2.4-6*x};
	%\addplot [domain=0:0.6,dashed,teal] {1.2-2*x}
	%	coordinate [pos=0.9] (r);
	\node (S1) at (axis cs:0.24,4*.24) [circle,fill,inner sep=1pt,label=0:$\text{S}_1$] {};
	\node (S2) at (axis cs:2.4/7,2.4/7) [circle,fill,inner sep=1pt,label=-20:$\text{S}_2$] {};
	
	\addplot [->,domain=0:0.2,samples=100,redorange](
		{0.34*exp(-167*x)*(-1.79*exp(2.36*x) + 1.37*exp(164.64*x) + exp(167*x))},
		{0.34*exp(-167*x)*(-0.02*exp(2.36*x) + 1.36*exp(164.64*x) + exp(167*x))})
		coordinate [pos=0] (t1a)
		coordinate [pos=0.5] (t1b);
	
	\addplot [->,domain=0:0.1,samples=100,redorange](
		{0.24*exp(-168*x)*(1.28*exp(3.37*x) - 0.61*exp(164.63*x) + exp(168*x))},
		{4*exp(-168*x)*(- 0.14*exp(164.63*x) + 0.24*exp(168*x))})
		coordinate [pos=0] (t2a)
		coordinate [pos=0.5] (t2b);
	
	\addplot [->,dashed,domain=0:0.4,samples=100,redorange](
		{(exp(-17*x)*(-4.99*exp(2.72*x) + 4.13*exp(14.28*x) + 2.06*exp(17.*x)))/6},
		{0.34*exp(-17.*x)*(-0.35*exp(2.72*x) + 1.68*exp(14.28*x) + exp(17.*x))});
	
	\addplot [->,dashed,domain=0:0.3,samples=100,redorange](
		{(exp(-18*x)*(2.6*exp(3.96*x) - 1.64*exp(14.04*x) + 1.44*exp(18*x)))/6},
		{4*exp(-18.*x)*(0.07*exp(3.96*x) - 0.21*exp(14.04*x) + 0.24*exp(18*x))});
	
	\node (A) at (t1a) [circle,fill,black,inner sep=1pt,label=180:A] {};
	\node (B) at (t2a) [circle,fill,black,inner sep=1pt,label=0:B] {};
	\node (A1) at (h2) [circle,fill,black,inner sep=1pt,label=135:A'] {};
	\node (B1) at (l2) [circle,fill,black,inner sep=1pt,label=-20:B'] {};
	%\draw [->] (A) to [bend left] (h2);
	%\draw [->] (B) to [bend left] (l2);
	\node at (axis cs:0.1,0.1) {I};
	\node at (axis cs:0.3,0.1) {N};
	\node at (axis cs:0.6,0.1) {M};
	
\nextgroupplot[
	xlabel={$u$},
	ylabel={$1-w$},
	y label style={at={(axis description cs:-0.01,.5)},anchor=south},
	title=(b) FHN Neuron]
	\xshade{axis cs:0,0}{axis cs:0.2,0}{red};
	\xshade{axis cs:0.2,0}{axis cs:0.4,0}{blue};
	\xshade{axis cs:0.4,0}{axis cs:1.1,0}{green};
	\draw (axis cs:0,0) -- (axis cs:0.2,0.8)
	-- (axis cs:0.4,0.4) -- (axis cs:1,1);
	%\addplot[dashed,thick,domain=0:1,red,samples=100] {62.5*x^3-52.5*x^2+11.9*x};
	\draw[dashed] (axis cs:0,0) -- (axis cs:0.4,0.4);
	\draw[dashed] (axis cs:0.2,0.8) -- (axis cs:0.3,1.2);
	\node at (axis cs:0.1,0.1) {I};
	\node at (axis cs:0.3,0.1) {N};
	\node at (axis cs:0.6,0.1) {M};
	\draw (axis cs:0.19,1.1) -- (axis cs:0.41,0);
	%\node[circle,fill,inner sep=1pt,label=45:S$_{b}$] at (axis cs:0.283,0.633) {};
	\node[circle,fill,inner sep=1pt,label=180:S$_{1}$] at (axis cs:0.228,0.911) {};
	\node[circle,fill,inner sep=1pt,label=-10:S$_{2}$] at (axis cs:0.342,0.342) {};
	\draw[->] (axis cs:0.32,0.8) -- (axis cs:0.38,0.8);
	\draw[->] (axis cs:0.3,0.85) -- (axis cs:0.38,0.85);
	\draw[->] (axis cs:0.28,0.9) -- (axis cs:0.38,0.9);
	\draw[->] (axis cs:0.3,0.4) -- (axis cs:0.22,0.4);
	\draw[->] (axis cs:0.28,0.45) -- (axis cs:0.22,0.45);
	\draw[->] (axis cs:0.26,0.5) -- (axis cs:0.22,0.5);

\end{groupplot}
\end{tikzpicture}
	\caption{(a) Trajectories (red) of system (\ref{eq:vo2_system}) in 
		the phase space $i_{i} \times (v_{dd}-v_{o})$ for a small $L/C$ value 
		(solid) and a large $L/C$ value (dashed). The 
		$i_{i}$-nullclines of system (\ref{eq:vo2_system}) are shown 
		as solid black lines in the metallic (AB') and insulating 
		(BA') states of the IMT device, and S$_{1}$S$_{2}$ is the 
		$v_{o}$-nullcline. Depending on the state, the phase space is divided into three 
		vertical regions - I, N and M. In the region N the $i_{i}$-nullclines are dependent on $s$ 
		(b) Nullclines of the FHN model in the phase space $u \times (1-w)$ where $f(u)$ is a piecewise linear function. The dynamics of FHN neuron are equivalent to the IMT neuron in the regions M and I. In the region N, for small $L/C$, the difference is only in the velocity and not the direction of system trajectories as they are parallel to $x$-axis}
	\label{fig:imt_traj}
\end{figure*}

\begin{figure}
	\centering
	\tikzsetnextfilename{imt_neuron_mosfet}
	\begin{tikzpicture}
\renewcommand{\baselinestretch}{1}
\begin{scope}[x=0.4cm,y=0.5cm]
	\draw
		(0,0) node[ground]{}
		to[short,i<^=$i_{o}$] ++(0,0.5) node(nmos)[nmos,anchor=S,scale=0.8]{}
		(nmos.G) to[short,-o] ++(-0.5,0) node[anchor=south]{$v_{gs}$}
		(nmos.D) -- (0,3) node(out){}
		to[L=$L$,i^<=$i_{i}$] ++(0,3)
		to[R=$g_{v}$] ++(0,1.5) 
		-- ++(0,0.5) node[vdd]{$v_{dd}$}
		++(0,-4.5) to[short,*-] ++(1.5,0)
		to[C,l_=$C$] ++(0,4.2) to[short,-*] ++(-1.5,0) ++(0,-4.7)
		to[short,*-o] ++(1.5,0) node[anchor=west]{$v_o$};
	\node at (-3,8.8) {(a)};
\end{scope}
\begin{scope}[shift={(5,0.2)}]
	\begin{axis}[
		xmin=0,
		xmax=0.6,
		ymin=0,
		ymax=1.3,
		ytick={0,0.4,0.8},
		yticklabels={0,$v_l$,$v_h$},
		xtick=\empty,
		xlabel={$i_{i}$},
		ylabel={$v_{dd}-v_{o}$},
		width=2.2in,
		height=2.1in,
		title=(b),
		grid=major]

		\yshade{axis cs:0,0.4}{axis cs:0,0.8}{teal};
		\addplot [domain=0:0.2] {4*x}
			coordinate [pos=0.5] (l2)
			coordinate [pos=1] (h1);
		\addplot [domain=0.4:1.2] {x}
			coordinate [pos=0] (l1)
			coordinate [pos=0.5] (h2);
		\addplot [domain=0.2:0.4,dashed] {4*x};
		\addplot [domain=0:0.4,dashed] {x};
		
		\addplot [domain=0:0.24,teal] 
			({1.2/3/0.24^2*(0.24-x/2)*x*(1+x/1.2)},{1.2-x});
		\addplot [domain=0.24:0.4,teal] {2.4-6*x}
			node [pos=0.4,pin={[black]45:Spiking}] {};
		
		\addplot [domain=0:0.2,redorange] 
			({1.2/3/0.24^2*(0.2-x/2)*x*(1+x/1.2)},{1.2-x});
		\addplot [domain=0.2:1.4,redorange] 
			({1.2/3/.24^2*0.2^2/2*(1+x/1.2)},{1.2-x})
			node [pos=0.2,circle,fill,inner sep=1pt] {}
			node [pos=0.627,circle,fill,inner sep=1pt] {}
			node [pos=0.4,pin={[black]260:Resting}] {};
		
		\node (S1) at (axis cs:0.24,4*.24) 
			[circle,fill,teal,inner sep=1pt,label=0:$\text{S}_1$] {};
		\node (S2) at (axis cs:2.4/7,2.4/7) 
			[circle,fill,teal,inner sep=1pt,label=-20:$\text{S}_2$] {};
		%\draw [->] (A) to [bend left] (h2);
		%\draw [->] (B) to [bend left] (l2);
	\end{axis}
\end{scope}
\end{tikzpicture}
	\caption{(a) IMT neuron with series transistor used to achieve bifurcation
		between a spiking and a resting state. (b) Nullclines of the system
		with series transistor in the phase space $i_{i} \times v_{dd}-v_{o}$ for
		two different $v_{gs}$ values for spiking and resting states. Bifurcation
		occurs when a stable points crosses the boundary of region $v_{dd}-v_{o}\in[v_{l},v_{h}]$.}
	\label{fig:imt_neuron_mosfet}
\end{figure}
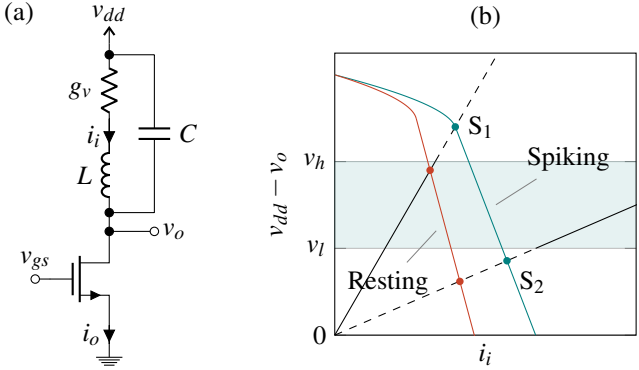

The model of (\ref{eq:vo2_system}) is very similar to a piecewise
linear caricature of FitzHugh-Nagumo (FHN) neuron model \citep{gerstner_spiking_2002},
also called the McKean's caricature \citep{tonnelier_mckeans_2003,mckean_nagumos_1970}.
Mathematically, the FHN model is given by:
\begin{align}
	\frac{du}{dt} & =f(u)-w+I_{ext}\nonumber \\
	\tau\frac{dw}{dt} & =u-bw+a\label{eq:fitzhugh_system}
\end{align}
where $f(u)$ is a polynomial of third degree, e.g. $f(u)=u-u^{3}/3$,
and $I_{ext}$ is the parameter for bifurcation, as opposed to $g_{s}$
in (\ref{eq:vo2_system}). In the FHN model, one variable ($u$),
possessing cubic nonlinearity, allows regenerative self-excitation
via a positive feedback, and the second, a recovery variable ($w$),
possessing linear dynamics, provides a slower negative feedback. It
was reasoned in Ref. \citep{mckean_nagumos_1970} that the 
essential features of FHN model are retained in a ``caricature'' where 
the cubic non-linearity is replaced by a piecewise linear function 
$f(u)$. Nullclines of (\ref{eq:fitzhugh_system}) with a piecewise linear $f(u)$ are shown in figure \ref{fig:imt_traj}b in the 
phase space $u \times (1-w)$. A function $f(u)$ is trivially possible such 
that it is equal to $v_{dd}-h(i_{i},s)$ in the regions M and I, hence making the $u$-nullcline similar to the $i_{i}$-nullcline in those regions. In the region N, the difference between $f(u)$ and $v_{dd}-h(i_{i},s)$ 
for any state $s$ does not result in a difference in the direction of 
system trajectories but only in their velocity, because for small 
$L/C$ the trajectories are almost parallel to $x$-axis. Bifurcation in VO$_{2}$ neuron is achieved by tuning the load line
using a tunable resistance ($g_{s}$), or a series transistor (figure
\ref{fig:imt_neuron_mosfet}a). Figure \ref{fig:imt_neuron_mosfet}b
shows two load line curves corresponding to different gate voltages
($v_{gs}$), where one gives rise to spikes while the other results
in a resting state.

\subsubsection{Single dimensional approximation}

\begin{figure}
	\centering
	\tikzsetnextfilename{exp_waves}
	\begin{tikzpicture} 
	\renewcommand{\baselinestretch}{1}
	\begin{groupplot}[
		group style={
			group size=1 by 3,
			xlabels at=edge bottom,
			xticklabels at=edge bottom,
			ylabels at=edge left,
			yticklabels at=edge left,
			vertical sep=0pt,
		},
		xlabel=Time ($\mu s$),
		ylabel=$v_{o}$ (V),
		minor x tick num=3,
		minor y tick num=1,
		width=3in,
		height=1.3in,
		ymin=0,
		xmin=0,
		xmax=500,
		ymax=6,
		ytick={0,2,4},
		grid=major,
		legend style={draw=none}]
	
		\nextgroupplot[
			legend pos=outer north east,
			legend entries={1.78V}]
			\addplot [mark=none,redorange] table {"\plotdatadir/exp_waveform_178.dat"};
		
		\nextgroupplot[
			legend pos=outer north east,
			legend entries={1.79V}]
			\addplot [mark=none,teal] table {"\plotdatadir/exp_waveform_179.dat"};
		
		\nextgroupplot[
			legend pos=outer north east,
			legend entries={1.81V}]
			\addplot [mark=none,cedar] table {"\plotdatadir/exp_waveform_181.dat"};
		
	\end{groupplot}
\end{tikzpicture}
	\caption{Experimental waveforms of VO$_{2}$ based spiking neuron for various
		$v_{gs}$ values (1.78V, 1.79V and 1.81V). A VO$_{2}$ neuron shows
		almost instantaneous charging (spike) in metallic state.}
	\label{fig:exp_waves}
\end{figure}
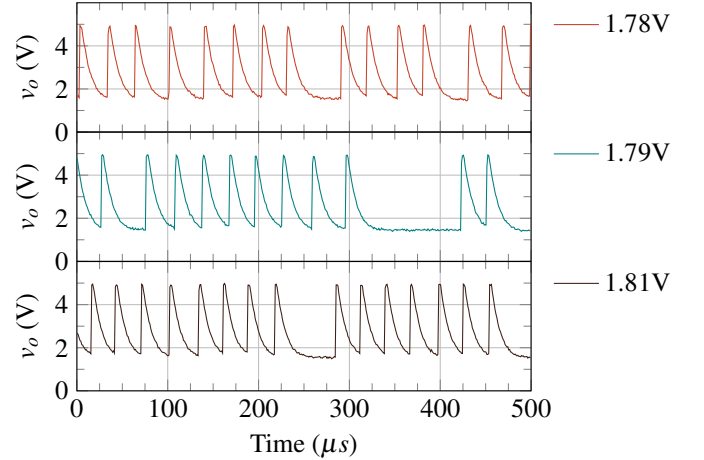

Moreover, a single dimensional piecewise approximation of the system
can be performed using a dimensionality reduction by replacing the
movement along the eigenvector parallel to the x-axis with an instantaneous
transition from A to A', or B to B'. This leaves a 1-dimensional subsystem
in M and I each along the V-I curves AB' and BA'. Experiments using VO$_{2}$ 
show that the metallic state conductance $g_{vm}$ is very high which causes
the charging cycle of $v_{o}$ to be almost instantaneous (figure
\ref{fig:exp_waves}) and resembles a spike of a biological neuron.
As such, the spiking statistics can be studied by modeling just the
discharge cycle of $v_{o}$. The inductance being negligible
can be effectively removed and only the capacitance is needed for
modeling the 1D subsystem of insulating state (figure \ref{fig:firing_rate}a)
making $v_{i}=v_{dd}-v_{o}$.

\subsection{Noise induced stochastic behavior}

\begin{figure*}
	\centering
	\tikzsetnextfilename{firing_rate}
	\begin{tikzpicture}
\renewcommand{\baselinestretch}{1}
\begin{scope}
	\tikzset{align at top/.style={baseline=(current bounding box.north)}}
	\draw
		(0,0) node[ground]{}
		to ++(0,0.5) node(nmos)[nmos,anchor=S]{}
		(nmos.G) to[short,-o] ++(-0.5,0) node[anchor=south]{$v_{gs}$}
		(nmos.D) to[voltage source=$\eta(t)$,invert,/tikz/circuitikz/bipoles/length=0.9cm] (0,4.5) node(out){}
		to ++(0,0.5)
		to[R=$g_{v(m/i)}$] ++(0,3) node[vdd]{$v_{dd}$}
		++(0,-3) -- ++(2,0)
		to[C,l_=$C$] ++(0,3) -- ++(-2,0)
		(out) to[short,*-o] ++(2.5,0) node[anchor=west]{$v_o$};
	\node at (-2,9.2) {(a)};
\end{scope}
\begin{scope}[shift={(7,0.9)}]
	\node at (-1,8.3) {(b)};
	\begin{groupplot}[
		group style={
			group size=3 by 1,
			xlabels at=edge bottom,
			xticklabels at=edge bottom,
			ylabels at=edge left,
			yticklabels at=edge left,
			horizontal sep=5pt},
		xlabel=Gate voltage $v_{gs}$ (V),
		ylabel=Firing rate (KHz),
		grid=major,
		width=2.1in,
		height=2in,
		ytick={0,10,20,30,40},
		legend style={draw=none},
		clip=false,
		y filter/.code={\pgfmathparse{#1*10^3}\pgfmathresult}
		]

	\nextgroupplot[
		legend entries={$\sigma_t=6$, $\sigma_t=7$, $\sigma_t=8$, VO$_2$},
		legend columns=4,
		legend style={at={(1,1.25)},anchor=south}]
		\draw [decorate,decoration={brace,amplitude=7pt}] 
			(rel axis cs:0.5,1.15) -- (rel axis cs:1.5,1.15);
		\addplot[mark=*,teal] table {"\plotdatadir/firing_rate_thermal_60.dat"};
		%\addplot+[redorange] table {"\plotdatadir/firing_rate_thermal_65.dat"};
		\addplot[mark=square*,redorange] table {"\plotdatadir/firing_rate_thermal_70.dat"};
		\addplot[mark=triangle*,cedar] table {"\plotdatadir/firing_rate_thermal_80.dat"};
		\addplot[mark=diamond*,moss] table {"\plotdatadir/firing_rate_exp.dat"};
		\node[anchor=west] at (rel axis cs:0.1,0.9) {$\mathbf{v_h} \sim $ Constant};

	\nextgroupplot
		\addplot[mark=*,teal] table {"\plotdatadir/firing_rate_both_60.dat"};
		%\addplot+[] table {"\plotdatadir/firing_rate_both_65.dat"};
		\addplot[mark=square*,redorange] table {"\plotdatadir/firing_rate_both_70.dat"};
		\addplot[mark=triangle*,cedar] table {"\plotdatadir/firing_rate_both_80.dat"};
		\addplot[mark=diamond*,moss] table {"\plotdatadir/firing_rate_exp.dat"};
		\node[anchor=west] at (rel axis cs:0.1,0.9) {$\mathbf{v_h} \sim $ Gaussian};

	\nextgroupplot[
		legend entries={$\sigma_t=4$, $\sigma_t=5$, $\sigma_t=6$, VO$_2$},
		legend columns=2,
		legend style={at={(0.5,1.15)},anchor=south}]
		\addplot[mark=*,teal] table 
			{"\plotdatadir/firing_rate_both_ep3_40.dat"};
		\addplot[mark=square*,redorange] table
			{"\plotdatadir/firing_rate_both_ep3_50.dat"};
		\addplot[mark=triangle*,cedar] table 
			{"\plotdatadir/firing_rate_both_ep3_60.dat"};
		\addplot[mark=diamond*,moss] table 
			{"\plotdatadir/firing_rate_exp.dat"};
		\node[anchor=west] at (rel axis cs:0.1,0.9) {$\mathbf{v_h} \sim $ EP[3]};

	\end{groupplot}
\end{scope}
\end{tikzpicture}
	\caption{(a) Noise model of IMT neuron where the noise components are the thermal
		noise voltage source $\eta(t)$ and the IMT threshold fluctuation.
		(b) Firing rate plotted against $v_{gs}$ using the analytical model
		for different $\mathbf{v_{h}}$ distributions (Constant, Gaussian,
		and EP{[}3{]}) and comparison with experimental observations.}
	\label{fig:firing_rate}
\end{figure*}
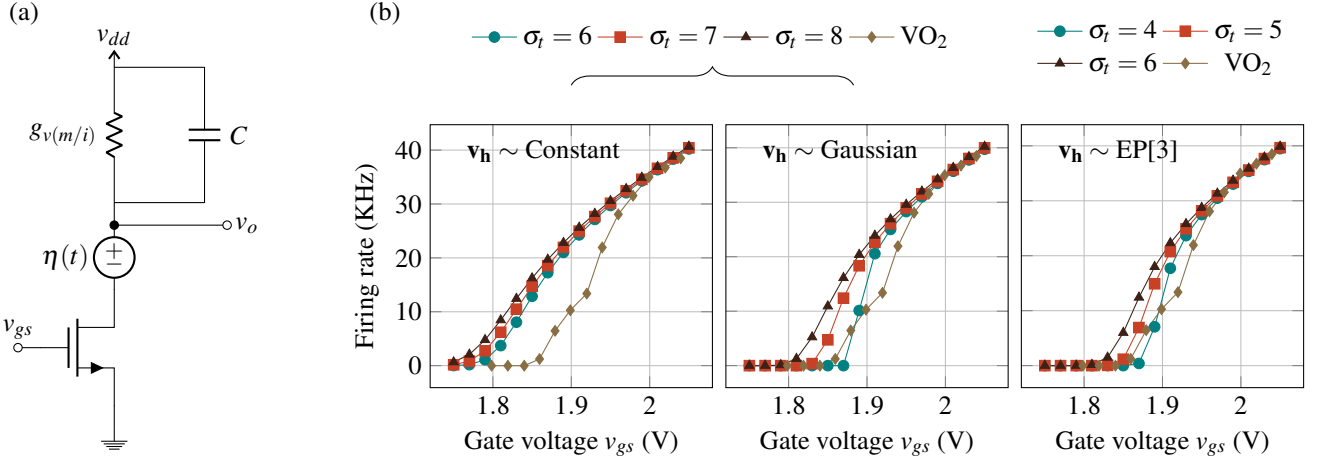

The two important noise sources which induce stochasticity in an IMT
neuron are (a) V$_{IMT}$ ($v_{h}$) fluctuations \citep{jerry2017random,zhang2016vo}, 
and (b) thermal noise. Thermal noise $\eta(t)$ is modeled in the circuit (figure
\ref{fig:firing_rate}a) as a white noise voltage $\eta(t)\mathrm{d}t=\sigma_{t}\mathrm{d}w_{t}$
where $w_{t}$ is the standard weiner process and $\sigma_{t}^{2}$
is the infinitesimal thermal noise variance. The threshold $v_{h}$
is assumed constant during a spike, but varies from one spike to another.
The distribution of $v_{h}$ from spike to spike is assumed to be
Gaussian or subGaussian whose parameters are estimated from experimental
observations of oscillations. If the series transistor always remains
in saturation and show linear voltage-current relationship, as is
the case in our VO$_{2}$ based experiments, the discharge phase can
be described by an Ornstein-Uhlenbeck (OU) process
\begin{equation}
	\mathrm{d}x=\frac{1}{\theta}(\mu-x)\mathrm{d}t+\sigma\mathrm{d}w_{t}\label{eq:ou_system}
\end{equation}
where $\mu$,$\theta$ and $\sigma$ are functions of circuit parameters
of the series transistor, the IMT device and $\sigma_{t}$. The interspike
interval is thus the first-passage-time (FPT) of this OU process,
but with a fluctuating boundary.

\subsubsection{OU process with constant boundary}

Analytical expressions for the FPT of OU process (with $\mu=0$) for
a constant boundary were derived using the Laplace transform method
in Ref. \citep{ricciardi_first-passage-time_1988}. Reproducing
some of its results, let the first passage time for the system (\ref{eq:ou_system}),
with $\mu=0$, which starts at $x(0)=x_{0}$ and hits a boundary $S$,
be denoted by the random variable $\mathbf{t_{f}}(S,x_{0})$, and
its $m^{th}$ moment by $\tau_{m}(S,x_{0})$. Also, let $\mathbf{\widetilde{t_{f}}}(S,x_{0})$
be the FPT for another OU process with $\mu=0$, $\theta=1$ and $\sigma=2$,
and $\widetilde{\tau_{m}}(S,x_{0})$ be its $m^{th}$ moment. Then
time and space scaling for the OU process imply that
\begin{align}
	\mathbf{t_{f}}(S,x_{0}) & \overset{d}{=}\theta\mathbf{\widetilde{t_{f}}}(\alpha S,\alpha x_{0})\nonumber \\
	\therefore\tau_{m}(S,x_{0}) & =\theta^{m}\widetilde{\tau_{m}}(\alpha S,\alpha x_{0})\label{eq:fpt_const}
\end{align}
where $\alpha=\sqrt{\frac{2}{\theta\sigma^{2}}}$. The first 2 moments
for the base case OU process $\widetilde{\tau_{1}}$ and $\widetilde{\tau_{2}}$
are given by
\begin{align}
	\widetilde{\tau_{1}}(S,x_{0})= & \phi_{1}(S)-\phi_{1}(x_{0})\nonumber \\
	\widetilde{\tau_{2}}(S,x_{0})= & 2\phi_{1}(S)^{2}-\phi_{2}(S) -2\phi_{1}(S)\phi_{1}(x_{0})+\phi_{2}(x_{0})\label{eq:tau_expansion}
\end{align}
where $\phi_{k}(z)$ can be written as an infinite sum
\begin{equation}
	\phi_{k}(z)=\frac{1}{2^{k}}\sum_{n=1}^{\infty}\frac{\left(\sqrt{2}z\right)^{n}\Gamma\left(\frac{n}{2}\right)\rho(n,k)}{n!}\label{eq:phi}
\end{equation}
with $\rho(n,k)$ being a function of the digamma function 
\citep{ricciardi_first-passage-time_1988}.

\subsubsection{OU process with fluctuating boundary}

We extend this framework for calculating the FPT statistics with a fluctuating
boundary $\mathbf{S}$ as follows. Let the IMT threshold be represented by the random
variable $\mathbf{v_{h}}$. For the VO$_{2}$ based IMT neuron, the
1D subsystem in the insulating phase can be converted in the form
of (\ref{eq:ou_system}) with $\mu=0$ by translating the origin to
the fixed point. If this transformation is T then $x=\textrm{T}v_{i}=\textrm{T}(v_{dd}-v_{o})$,
$\mathbf{S}=\text{T}\mathbf{v_{h}}$ and $x_{o}=\textrm{T}v_{l}$.
The start and end points are B' and A respectively in figure \ref{fig:imt_traj}.
$\mathbf{v_{h}}$ is assumed constant during a spike, and across spikes
the distribution of $\mathbf{v_{h}}$ is $\mathbf{v_{h}}\sim\mathcal{D}$,
where $\mathcal{D}$ is either Gaussian, or subGaussian. For subGaussian
distributions we use the Exponential Power family EP{[}$\kappa${]},
$\kappa$ being the shape factor. Let the interspike interval of IMT
neuron be denoted by the marginal random variable \textbf{$\mathbf{t_{imt}}(\mathcal{D},v_{l})$.}
Then $\mathbf{t_{imt}}$ is related to $\mathbf{t_{f}}$ in equation
(\ref{eq:fpt_const}), given common parameters $\theta$ and $\sigma$,
as follows:
\begin{align*}
	\mathbf{t_{imt}}(\mathcal{D},v_{l})|(\mathbf{v_{h}}=v) & \overset{d}{=}\mathbf{t_{f}}(\text{T}v,\text{T}v_{l})
\end{align*}
The moments of $\mathbf{t_{imt}}$ can be calculated as:
\begin{align}
	\mathbb{E}[\mathbf{t_{imt}}(\mathcal{D},v_{l})^{m}] & =\mathbb{E}_{v_{h}}[\mathbb{E}[\mathbf{t_{imt}}(\mathcal{D},\text{T}v_{l})^{m}|\mathbf{v_{h}}=v]]\nonumber \\
	& =\mathbb{E}_{v_{h}}[\tau_{m}(\text{T}\mathbf{v_{h}},\text{T}v_{l})]\nonumber \\
	& =\theta^{m}\mathbb{E}_{v_{h}}[\widetilde{\tau_{m}}(\alpha\text{T}\mathbf{v_{h}},\alpha\text{T}v_{l})]\label{eq:expectation_tower}
\end{align}
where $\alpha=\sqrt{\frac{2}{\theta\sigma^{2}}}$. If $\mathcal{D}$
is Gaussian or EP[$\kappa$] distribution and $\alpha\textrm{T}$
is an affine transformation, then $\alpha\textrm{T}\mathbf{v_{h}}$
also has a Gaussian or EP[$\kappa$] distribution.

\subsection{Experiments}

IMT devices are fabricated on a $10$nm VO$_{2}$ thin film grown by reactive
oxide molecular beam epitaxy on (001) TiO$_{2}$ substrate using a Veeco Gen10
system \citep{tashman2014epitaxial}. Planar two terminal structures are formed by
patterning contacts using standard electron beam lithography which defines the
device length (L$_{VO2}$). Pd ($20$nm) / Au ($60$nm) contacts are then deposited
by electron beam evaporation and liftoff. The devices are then isolated and the
widths (W$_{VO2}$) are defined using a CF$_{4}$ based dry etch.

\begin{figure*}
	\centering
	\tikzsetnextfilename{exp_figures}
	\begin{tikzpicture}
\begin{scope}
  \begin{axis}[
    xlabel=Current ($\mu$A),
    ylabel=Voltage (V),
    title=(a),
    width=2.5in,
    height=2.3in,
	every x tick scale label/.style={
	    at={(1,0)},xshift=1pt,anchor=south west,inner sep=0pt
	},
	grid=major,
	xmin=0,
	ymin=0,
	x filter/.code={\pgfmathparse{#1*10^6}\pgfmathresult}
  ]
    \foreach \x in {1,...,126}
    {\addplot[mark=none,lightgray] table[x=sw\x,y=V] {\plotdatadir/iv_curves.dat};}
    \addplot[mark=none,thick,redorange] table[x=sw10,y=V] {\plotdatadir/iv_curves.dat};
  \end{axis}
\end{scope}
\begin{scope}[shift={(10.7,0)}]
  \begin{axis}[
    xtick={1,2,3},
    xticklabels={116,175,215},
    boxplot/draw direction=y,
    xlabel=Peak Current ($\mu$A),
    ylabel=$\Delta V_{IMT}$ (V),
    title=(b),
    width=2.5in,
    height=2.3in,
    ymajorgrids
  ]
    \addplot+[boxplot] table[y index=0] {\plotdatadir/dvimt_116.dat};
    \addplot+[boxplot] table[y index=0] {\plotdatadir/dvimt_175.dat};
    \addplot+[boxplot] table[y index=0] {\plotdatadir/dvimt_215.dat};
  \end{axis}
\end{scope}
\end{tikzpicture}
	\caption{(a) The prototypical DC voltage-current characteristics for a single VO$_2$ device exhibits abrupt threshold switching at V$_{IMT}$ and V$_{MIT}$. The current in the metallic state has been arbitrarily limited to a 200\textmu A compliance current. (b) V$_{IMT}$ distribution as a function of the peak current during oscillations (value is set by the MOSFET saturation current). V$_{IMT}$ is extracted from 300+ cycles.}
	\label{fig:exp_iv}
\end{figure*}
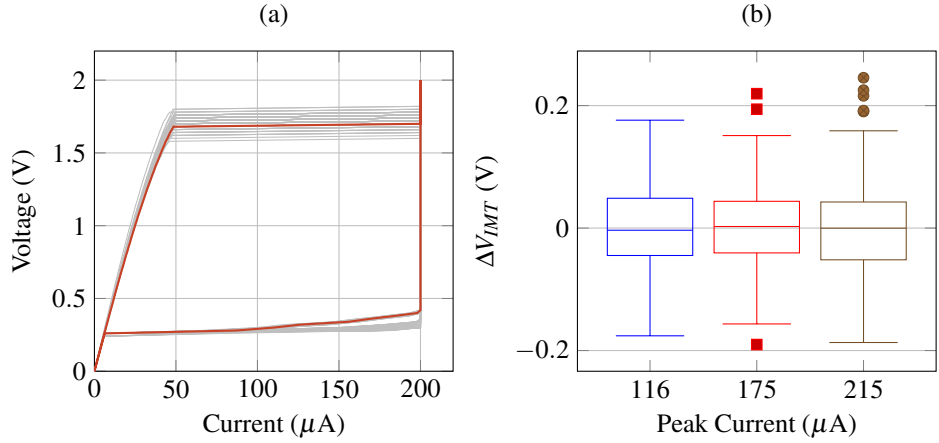

The IMT neuron is constructed using an externally connected n-channel MOSFET
(ALD110802) and the fabricated VO$_{2}$ device. A prototypical I-V curve is
shown in figure \ref{fig:exp_iv}a. Within the experimental data, the current is
limited to an arbitrarily chosen 200 $\mu$A to prevent a thermal runaway and
breakdown of the device while in the low resistance metallic state. It should be
noted that as the metallic state corresponds to the abrupt charging cycle of
$v_{o}$, limiting the current would not have noticeable effect on spiking
statistics of the neuron.

Threshold voltage fluctuations (cycle to cycle) were observed in all devices
which were tested ($>10$). Threshold voltage distribution was estimated using
the varying cycle-to-cycle threshold voltages collected from a single device.
Thermal noise is not measured directly, but is estimated approximately by
matching the simulation waveforms from the circuit model (Figure
\ref{fig:firing_rate}a) with the observed experimental waveforms. It can be
verified that thermal noise of the transistor is not the dominant noise source
by measuring the threshold variation as a function of the transistor current
(Figure \ref{fig:exp_iv}b) and observing that the distribution of switching
threshold does not change with varying transistor current. Finally, the firing
rate and its variation with $v_{gs}$ (Figure \ref{fig:firing_rate}b) were
measured for a single device.

\section{Results}

\subsection{Spiking Statistics}

\subsubsection{First moment and the firing rate}

First moment of $\mathbf{t_{imt}}$ is calculated using (\ref{eq:tau_expansion})
and (\ref{eq:expectation_tower}) as
\begin{align*}
	\mathbb{E}[\mathbf{t_{imt}}(\mathcal{D},v_{l})] & =\theta (\mathbb{E}_{v_{h}}[\phi_{1}(\alpha\text{T}\mathbf{v_{h}})]-\phi_{1}(\alpha x_{0}))
\end{align*}
The expansion for $\phi_{k}(z)$ in (\ref{eq:phi}) can be used to
calculate $\mathbb{E}_{v_{h}}[\phi_{k}(\alpha\textrm{T}\mathbf{v_{h}})]$
using the moments of $\alpha\textrm{T}\mathbf{v_{h}}$ as follows
\[
\mathbb{E}_{v_{h}}[\phi_{k}(\alpha\textrm{T}\mathbf{v_{h}})]=\frac{1}{2^{k}}\sum_{n=1}^{\infty}\frac{(\sqrt{2})^{n}\mathbb{E}[(\alpha\textrm{T}\mathbf{v_{h}})^{n}]\Gamma\left(\frac{n}{2}\right)\rho(n,k)}{n!}
\]
Figure \ref{fig:firing_rate}b shows firing rate ($1/\mathbb{E}[\mathbf{t_{imt}}(\mathcal{D},v_{l})]$)
as a function of $v_{gs}$ for various $\sigma_{t}$ values and for
3 distrbutions of threshold fluctuations. The calculations approximate
the experimental observations well for all three $v_{h}$ distributions,
the closest being EP{[}3{]} with $\sigma_{t}=4$.

\subsubsection{Higher moments}

For higher moments, higher order terms are encountered. For example,
in case of the second moment, using (\ref{eq:tau_expansion}) and (\ref{eq:expectation_tower}),
we obtain
\begin{align*}
	\mathbb{E}_{v_{h}}[\widetilde{\tau_{2}}(\alpha\textrm{T}\mathbf{v_{h}},\alpha\text{T}v_{l})]= & 2\mathbb{E}_{v_{h}}[\phi_{1}(\alpha\textrm{T}\mathbf{v_{h}})^{2}]-\mathbb{E}_{v_{h}}[\phi_{2}(\alpha\textrm{T}\mathbf{v_{h}})]\\
	& -2\mathbb{E}_{v_{h}}[\phi_{1}(\alpha\textrm{T}\mathbf{v_{h}})]\phi_{1}(\alpha\text{T}v_{l})\\
	& +\phi_{2}(\alpha\text{T}v_{l})
\end{align*}
with a higher order term $\phi_{1}(\alpha\textrm{T}\mathbf{v_{h}}){}^{2}$. 
In the case of the third moment we obtain $\phi_{1}(\alpha\textrm{T}\mathbf{v_{h}})\phi_{2}(\alpha\textrm{T}\mathbf{v_{h}})$.
As each $\phi_{k}$ term is an infinite sum, we construct a cauchy
product expansion for the higher order term using the infinite sum
expansions of the constituent $\phi_{k}$s and then distribute the
expectation over addition. For example, if the $\phi_{k}$ expansions
of $\phi_{1}(z)$ and $\phi_{2}(z)$ are $(\sum a_{i})$ and $(\sum b_{i})$
respectively, then the cauchy product expansion of $\phi_{1}(z)\phi_{2}(z)$
can be calculated as $\sum c_{i}$, where $c_{i}$ is a function of
$a_{1...i}$ and \textbf{$b_{1...i}$}, and the expectation $\mathbb{E}[\phi_{1}(z)\phi_{2}(z)]=\sum\mathbb{E}[c_{i}]$.
Since $c_{i}$ is a polynomial in $z$, $\mathbb{E}[c_{i}]$ can be calculated
using the moments of $z$.

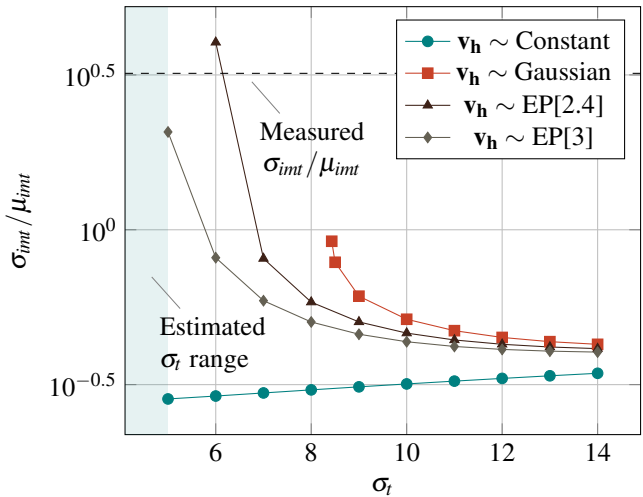
\begin{figure}
	\centering
	\tikzsetnextfilename{variance}
	\begin{tikzpicture} 
	\renewcommand{\baselinestretch}{1}
	\begin{semilogyaxis}[
		xlabel=$\sigma_t$,
		ylabel=$\sigma_{imt}/\mu_{imt}$,
		grid=major,
		legend entries={$\mathbf{v_{h}} \sim $ Constant, $\mathbf{v_{h}} \sim $ Gaussian, $\mathbf{v_{h}} \sim $ EP[2.4], $\mathbf{v_{h}} \sim $ EP[3]},
		legend pos=north east,
		restrict y to domain =-2:3]

		\addplot [mark=*,teal] table 
			{"\plotdatadir/coeff_var_thermal.dat"}
			coordinate [pos=0] (A);
		\addplot [mark=square*,redorange] table 
			{"\plotdatadir/coeff_var_both_normal.dat"}
			coordinate [pos=0] (B);
		\addplot [mark=triangle*,cedar] table 
			{"\plotdatadir/coeff_var_both_ep24.dat"};
		\addplot [mark=diamond*,redorange!50!teal] table
			{"\plotdatadir/coeff_var_both_ep3.dat"};
		%\draw [<->,shorten >=5,shorten <=5] (A) -- (B)
		%	node[pos=0.25,anchor=west]{$3\times$};
		\xshade {axis cs:3,0}{axis cs:5,0}{teal};
		\node (E) [pin={[align=center]280:Measured\\ $\sigma_{imt}/\mu_{imt}$}] 
			at (axis cs:6.6,3.2) {};
		\yline {E};
		\node [pin={[align=left]280:Estimated\\ $\sigma_{t}$ range}] 
			at (axis cs:4.5,0.75) {};
		%\addplot[mark=otimes,red,fill=red!20] coordinates {(4.9,3.2)};
\end{semilogyaxis}
\end{tikzpicture}
	\caption{$\sigma_{imt}/\mu_{imt}$ for the interspike interval plotted against
		$\sigma_{t}$ for $v_{gs}=1.8$V with Constant, Gaussian and Exponential
		Power (EP{[}$\kappa${]}, where $\kappa$ is the shape factor) distributions
		of the threshold noise. The experimentally observed $\sigma_{imt}/\mu_{imt}$
		for a VO$_{2}$ neuron is shown with a dotted line. The shaded region shows
		the experimentally estimated range of $\sigma_{t}$ ($\sigma_{t}<5$)}
	\label{fig:variance}
\end{figure}

If $\mu_{imt}$ and $\sigma_{imt}$ are the mean and standard deviation
of interspike intervals $\mathbf{t_{imt}}$, the coefficient of variation ($\sigma_{imt}/\mu_{imt}$)
varies with the relative proportion of the thermal and the threshold induced noise.
Figure \ref{fig:variance} shows $\sigma_{imt}/\mu_{imt}$ (calculated
using parameters matched with our VO$_{2}$ experiments) plotted against
$\sigma_{t}$ for various kinds of $\mathbf{v_{h}}$ distributions
fitted to experimental observations. $\sigma_{imt}/\mu_{imt}$ as
observed in our VO$_{2}$ experiments is about an order of magnitude
more than what would be calculated with only thermal noise using such
a neuron, and hence, threshold noise contributes significant stochasticity
to the spiking behavior. As the IMT neuron is setup such that the
stable point is close to the IMT transition point (figure \ref{fig:imt_neuron_mosfet}b),
low $\sigma_{t}$ results in high and diverging $\sigma_{imt}/\mu_{imt}$
for any distribution of threshold noise, and $\sigma_{imt}/\mu_{imt}$ reduces
with increasing $\sigma_{t}$ for the range shown. For a Normally distributed $v_{h}$
the variance diverges for $\sigma_{t}\lesssim8$, but for Exponential
Power (EP) distributions with lighter tails, the variance converges
for smaller values of $\sigma_{t}$. Statistical measurements on experimental 
data, as indicated in figure \ref{fig:variance}, provide measures of 
$\sigma_{imt}/\mu_{imt}$ (dotted line) and $\sigma_{t}$ (shaded region). We 
note that EP distributions provide a better approximation of the stochastic 
nature of experimentally demonstrated  VO$_{2}$ neurons as the range of 
$\sigma_{t}$ is estimated to be less than $5$.

\section{Discussion}

In this paper, we demonstrate and analyse an IMT based stochastic neuron
hardware which relies on both threshold fluctuations and thermal noise as
precursors to bifurcation. The IMT neuron emulates the functionality of
theoretical neuron models completely by incorporating all neuron characteristics
into device phenomena. Unlike other similar efforts, it does not need peripheral
circuits alongside the core device circuit (an IMT device and a transistor) to
emulate any sub-component of the spiking neuron model like thresholding, reset
etc. Moreover, the neuron construction not only utilizes inherent physical noise
sources for stochasticity, but also enables control of firing probability using
an analog electrical signal - the gate voltage of series transistor. This is
different from previous works which control only the deterministic aspect of
firing rate like the charging rate. A comparison of  spiking neuron hardware
characteristics in different works is shown in Table \ref{table:comparison}.

\begin{table*}
	\def\tabularxcolumn#1{m{#1}}
	\setlength{\extrarowheight}{3pt}
	\newcolumntype{P}{>{\raggedright\arraybackslash}X}
	\begin{tabularx}{\textwidth}{>{\bfseries}PPPPP>{\color{redorange}}P}
		\hline
		\rowcolor{gray!30}[2pt][2pt]
		 & \textbf{\textcite{tuma2016stochastic}} & \textbf{\textcite{pickett2013scalable}} & \textbf{\textcite{sengupta2016magnetic}} & \textbf{\textcite{indiveri2006vlsi}} & \textbf{This work (VO$_2$)}\tabularnewline
		\hline  
		Neuron type & Integrate \& Fire & Hodgkin Huxley & Integrate \& Fire & Integrate \& Fire & Piecewise Linear FHN\tabularnewline
		%\hline 
		Material / Platform & Chalcogenide & Mott insulator NbO$_2$ & MTJ  & 0.35 \textmu m CMOS & Vanadium Dioxide (VO$_2$)\tabularnewline
		%\hline 
		Material phenomenon & Phase Change  & IMT & Spin transfer torque (STT) & - & IMT\tabularnewline
		\hline 
		Spontaneous spiking using only device & No & Yes & No & - & Yes\tabularnewline
		\hline 
		Peripherals needed for spiking & Yes, for spike generation and reset & No & Yes, for spike generation and reset & - & No\tabularnewline
		%\hline 
		Integration mechanism (I\&F) & Heat accumulation & - & Magnetization accumulation & Capacitor charging & Capacitor charging\tabularnewline
		%\hline 
		Threshold mechanism (I\&F) & External reset by measuring conductance & Spontaneous IMT & External reset by detecting magnet flip & Reset using comparator & Spontaneous IMT\tabularnewline
		\hline 
		Stochastic & Yes & -  & Yes & No & Yes\tabularnewline
		\hline 
		Kind of stochasticity (I\&F) & Reset potential & - & Differential & - & Threshold and differential\tabularnewline
		%\hline 
		Source of stochasticity / noise & Melt-quench process & - & Thermal noise & - & IMT threshold fluctuations \& Thermal noise\tabularnewline
		%\hline 
		Control of stochastic firing rate & Only integration rate & - & Only integration rate & Only integration rate & Yes\tabularnewline
		\hline 
		Status of experiments & Constant stochasticity, variable integration rate & Deterministic spiking & None & Deterministic spiking & Sigmoidal variation of stochastic firing rates\tabularnewline
		\hline 
		Peak current & 750-800 \textmu A &  & - &  & 200 \textmu A \tabularnewline
		%\hline 
		Power or Energy/spike & 120 \textmu W &  & - & 900 pJ / spike & 196 pJ / spike\tabularnewline
		%\hline 
		Voltage & 5.5 V & 1.75 V & - & 3.3 V & 0.7V \tabularnewline
		%\hline 
		Maximum firing rates & 35-40 KHz & 30 KHz & - & 200 Hz & 30 KHz \tabularnewline
		\hline 
	\end{tabularx}
	\caption{Comparison of this work (experimental details from \textcite{Jerry2017}) with other spiking neuron hardware works based on different characteristics of spiking neurons}
	\label{table:comparison}
\end{table*}

We also show that the neuron dynamics follow a linear ``carricature'' of the
FitzHugh-Nagumo model with intrinsic stochasticity. The analytical models
developed in this paper can also faithfully reproduce the experimentally
observed transfer curve which is a stochastic property. Such analytical
verification of stochastic neuron experiments is one of the first in this work.
It is an important result as it indicates reproducibility of stochastic
characteristics and helps in creating the pathway towards perfecting these
devices. With a growing concensus that stochasticity will play a key role in
solving hard computing tasks, we need efficient ways for controlled
amplification and conversion of physical noise into a readable and computable
form. In this regard, the IMT based neuron represents a promising solution for a
stochastic computational element. Such stochastic neurons have the potential to
realize bio-mimetic computational kernels that can be employed to solve a large
class of optimization and machine-learning problems.

\section*{Acknowledgements}
This project was supported by the National Science Foundation under grants 
1640081, Expeditions in Computing Award-1317560 and CCF- 1317373, and the Nanoelectronics Research Corporation (NERC), a wholly-owned
subsidiary of the Semiconductor Research Corporation (SRC), through Extremely 
Energy Efficient Collective Electronics (EXCEL), an SRC-NRI Nanoelectronics 
Research Initiative under Research Task IDs 2698.001 and 2698.002.

\section*{Author Contributions}
A.P. worked on the development of theory, simulation frameworks and mathematical models. M.J. worked on the experiments. A.R. advised A.P. and participated in the problem formulation. S.D. advised M.J. and also participated in the design of experiments and problem formulations.

\section*{Conflict of Interest Statement}
%All financial, commercial or other relationships that might be perceived by the academic community as representing a potential conflict of interest must be disclosed. If no such relationship exists, authors will be asked to confirm the following statement: 
The authors declare that the research was conducted in the absence of any commercial or financial relationships that could be construed as a potential conflict of interest.

\bibliography{paper_refs}

\end{document}